\documentclass{article}      

\newcommand{\ba}{\begin{eqnarray}}
\newcommand{\ea}{\end{eqnarray}}
\begin{document}
\pagestyle{plain}

\title{Magnetic moments of antidecuplet pentaquarks}  
\author{R. Bijker\\
Instituto de Ciencias Nucleares, 
Universidad Nacional Aut\'onoma de M\'exico,\\
A.P. 70-543, 04510 M\'exico, D.F., M\'exico
\and
M.M. Giannini and E. Santopinto \\
Dipartimento di Fisica dell'Universit\`a di Genova, 
I.N.F.N., Sezione di Genova, \\
via Dodecaneso 33, 16164 Genova, Italy}
\date{May 5, 2004}
\maketitle                   

\begin{abstract}
We analyze the magnetic moments of the exotic pentaquarks of the flavor 
antidecuplet in the constituent quark model for the cases in which the 
ground state is in an orbital $L^p=0^+$ or a $L^p=1^-$ state. We derive 
a set of sum rules for the magnetic moments of antidecuplet baryons 
and their relation with the magnetic moments of decuplet and octet baryons. 
The magnetic moment of the $\Theta^+(1540)$ is found to be 0.38, 0.09 
and 1.05 $\mu_N$ for $J^p=1/2^-$, $1/2^+$ and $3/2^+$, respectively, 
which is compared with the results obtained in other approaches. 
\end{abstract}

\begin{center}
\end{center}

\section{Introduction}

The discovery of the $\Theta^+(1540)$ resonance with positive strangeness 
$S=+1$ by the LEPS Collaboration \cite{leps} and its subsequent 
confirmation by various other experimental collaborations \cite{penta} has 
motivated an enormous amount of experimental and theoretical studies of 
exotic baryons \cite{jlab}, despite some other reports 
in which the pentaquark signal is attributed to kinematical reflections 
from the decay of mesons \cite{dzierba}, or in which no evidence has been 
found for such states \cite{nopenta}. 
More recently, the NA49 Collaboration \cite{cern} reported evidence for the 
existence of another exotic baryon $\Xi^{--}(1862)$ with strangeness 
$S=-2$, although this claim has been shown to be, at least partially, 
inconsistent with the existing data on the spectroscopy of cascade baryons 
\cite{cascade}. Results from the CLAS Collaboration are inconclusive for  
lack of statistics \cite{nefkens}. The $\Theta^+$ and $\Xi^{--}$ resonances 
are interpreted as $q^4 \bar{q}$ pentaquarks belonging to a flavor 
antidecuplet with quark structure $uudd\bar{s}$ and $ddss\bar{u}$, 
respectively. In addition, there is now the first evidence \cite{h1} for a 
heavy pentaquark at 3099 MeV in which the antistrange quark in the 
$\Theta^+$ is replaced by an anticharm quark. 

The spin and parity of the $\Theta^+$ have not yet been determined 
experimentally. The parity of the pentaquark ground state is predicted to be 
positive by many studies, such as chiral soliton models \cite{soliton}, 
cluster models \cite{JW,SZ,KL}, lattice QCD \cite{chiu}, and various 
constituent quark models \cite{cqmpos}. 
However, there are also many predicitions for 
a negative parity ground state pentaquark from recent work on QCD sum rules 
\cite{sumrule}, lattice QCD \cite{lattice}, quark model calculations 
\cite{cqmneg,BGS1,BGS2}, as well as from a study in the chiral soliton 
model \cite{wuma}. Many different 
proposals have been made to measure the parity in nucleon-nucleon collisions 
\cite{thomas} and in photoproduction experiments \cite{photo,Nam,Zhao}. 

Another unknown quantity is the magnetic moment. Although it may be difficult 
to determine its value experimentally, it is an essential ingredient 
in calculations of the photo- and electroproduction cross sections 
\cite{photo,Nam,Zhao}. Meanwhile, in the absence of experimental information, 
one has to rely on model calculations. The magnetic moment of the $\Theta^+$ 
pentaquark has been calculated in a variety of approaches 
\cite{Nam,Zhao,Kim1,Huang,Liu,Li} ranging from correlated quark models, 
the chiral soliton model, QCD sum rules and the MIT bag model. It is of 
interest to compare these results with those for the constituent 
quark model as well, since different values of the magnetic moments have 
their consequences for the photoproduction cross sections. 
which in turn may be used to help determine the quantum numbers of the 
$\Theta^+$ \cite{Nam,Zhao}.

The aim of this Letter is to study the magnetic moment of antidecuplet 
pentaquarks in the constituent quark model. In general, the pentaquark 
spectrum contains states of both positive and negative parity. Their 
precise ordering, in particular the angular momentum and parity of the 
ground state, depends on the choice of a specific dynamical model 
and the relative size of the orbital excitations, the 
spin-flavor splittings and the spin-spin couplings 
\cite{cqmpos,cqmneg,BGS1,BGS2,dudek}. The present analysis is carried out 
for antidecuplet pentaquarks of both parities: $J^p=1/2^-$, $1/2^+$ and 
$3/2^+$. We derive a set of sum rules for the magnetic moments, and make a 
comparison with the results obtained in correlated (or cluster) quark models. 
Some preliminary results of this work have been published in \cite{BGS2}

\section{Pentaquark wave functions}

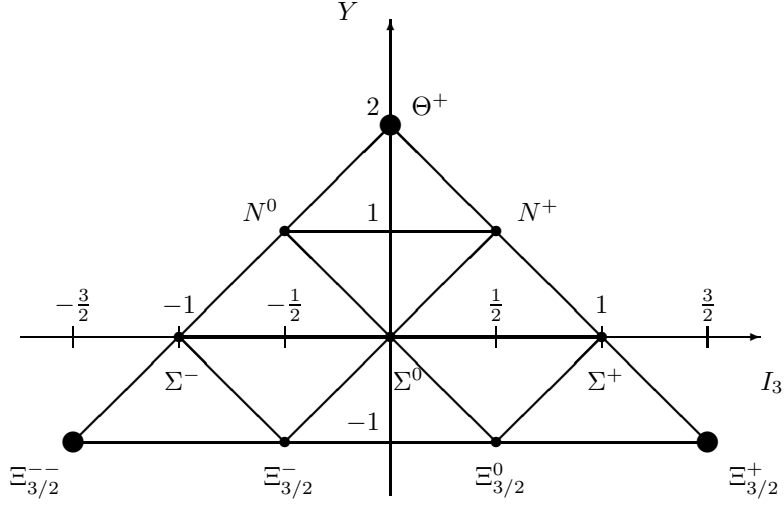
\begin{figure}
\centering
\setlength{\unitlength}{0.8pt}
\begin{picture}(350,200)(75,75)
\thinlines
\put( 75,150) {\vector(1,0){350}}
\put(250, 75) {\vector(0,1){225}}
\put(100,145) {\line(0,1){10}}
\put(150,145) {\line(0,1){10}}
\put(200,145) {\line(0,1){10}}
\put(250,145) {\line(0,1){10}}
\put(300,145) {\line(0,1){10}}
\put(350,145) {\line(0,1){10}}
\put(400,145) {\line(0,1){10}}
\thicklines
\put(200,200) {\line(1,0){100}}
\put(150,150) {\line(1,0){200}}
\put(100,100) {\line(1,0){300}}

\put(100,100) {\line(1,1){150}}
\put(200,100) {\line(1,1){100}}
\put(300,100) {\line(1,1){ 50}}

\put(200,100) {\line(-1,1){ 50}}
\put(300,100) {\line(-1,1){100}}
\put(400,100) {\line(-1,1){150}}

\multiput(250,250)(100,0){1}{\circle*{5}}
\multiput(200,200)(100,0){2}{\circle*{5}}
\multiput(150,150)(100,0){3}{\circle*{5}}
\multiput(100,100)(100,0){4}{\circle*{5}}

\multiput(100,100)(300,0){2}{\circle*{10}}
\put(250,250){\circle*{10}}

\put(260,255){$\Theta^+$}
\put(180,205){$N^0$}
\put(310,205){$N^+$}
\put(143,125){$\Sigma^-$}
\put(251,125){$\Sigma^0$}
\put(343,125){$\Sigma^+$}
\put( 70, 80){$\Xi_{3/2}^{--}$}
\put(190, 80){$\Xi_{3/2}^{-}$}
\put(290, 80){$\Xi_{3/2}^{0}$}
\put(410, 80){$\Xi_{3/2}^+$}

\put(425,125){$I_3$}
\put(225,300){$Y$}
\put(100,165){\makebox(0,0){$-\frac{3}{2}$}}
\put(150,165){\makebox(0,0){$-1$}}
\put(200,165){\makebox(0,0){$-\frac{1}{2}$}}
\put(300,165){\makebox(0,0){$ \frac{1}{2}$}}
\put(350,165){\makebox(0,0){$ 1$}}
\put(400,165){\makebox(0,0){$ \frac{3}{2}$}}
\put(245,255){\makebox(0,0)[br]{$2$}}
\put(245,205){\makebox(0,0)[br]{$1$}}
\put(245,105){\makebox(0,0)[br]{$-1$}}
\end{picture}
\caption[]{$SU(3)$ flavor multiplet $[33]$ with $E$ symmetry. 
The isospin-hypercharge multiplets are $(I,Y)=(0,2)$, $(\frac{1}{2},1)$, 
$(1,0)$ and $(\frac{3}{2},-1)$. Exotic states are indicated with $\bullet$.}
\label{flavor33e}
\end{figure}

We consider pentaquarks to be built of five constituent parts which are 
characterized by both internal and spatial degrees of freedom. 
The internal degrees of freedom are taken to be the three light flavors 
$u$, $d$, $s$ with spin $s=1/2$ and three colors $r$, $g$, $b$. 
The corresponding algebraic structure consists of the usual spin-flavor 
and color algebras $SU_{\rm sf}(6) \otimes SU_{\rm c}(3)$.
In the construction of the classification scheme we are guided by two 
conditions: the pentaquark wave function should be antisymmetric under 
any permutation of the four quarks, and should be a color singlet. 
The permutation symmetry of the four-quark subsystem is characterized by  
the $S_4$ Young tableaux $[4]$, $[31]$, $[22]$, $[211]$ and $[1111]$ or, 
equivalently, by the irreducible representations of the tetrahedral group  
${\cal T}_d$ (which is isomorphic to $S_4$) as $A_1$ (symmetric), $F_2$, 
$E$, $F_1$ (mixed symmetric) and $A_2$ (antisymmetric), respectively. 
For notational purposes we prefer to use the latter to label the 
discrete symmetry of the pentaquark wave functions. 
The corresponding dimensions are 1, 3, 2, 3 and 1, respectively.  
The full decomposition of the spin-flavor states into spin and 
flavor states $SU_{\rm sf}(6) \supset SU_{\rm f}(3) \otimes SU_{\rm s}(2)$ 
is given in Table~5 of \cite{BGS1}. The states of a given flavor multiplet 
can be labeled by isospin $I$, $I_3$ and hypercharge $Y$. 
It is difficult to distinguish the pentaquark flavor singlets, octets 
and decuplets from the three-quark flavor multiplets, since they have 
the same values of the hypercharge $Y$ and isospin projection $I_3$. The same 
observation holds for the majority of the states in the remaining flavor 
states. However, the antidecuplets, the 27-plets and 35-plets contain 
in addition exotic states which cannot be obtained from three-quark 
configurations. These states are more easily identified experimentally 
due to the uniqueness of their quantum numbers. 
The recently observed $\Theta^+$ and $\Xi^{--}$ resonances are 
interpreted as pentaquarks belonging to a flavor antidecuplet with 
isospin $I=0$ and $I=3/2$, respectively. In Fig.~\ref{flavor33e} 
the exotic states are indicated by a $\bullet$: 
the $\Theta^+$ is the isosinglet $I=I_3=0$ with hypercharge $Y=2$ 
(strangeness $S=+1$), and the cascades $\Xi_{3/2}^+$ and $\Xi_{3/2}^{--}$ 
have hypercharge $Y=-1$ (strangeness $S=-2$) and isospin $I=3/2$ with 
projection $I_3=3/2$ and $-3/2$, respectively. 

A convenient choice to describe the relative motion of the constituent 
parts is provided by the Jacobi coordinates \cite{KM}
\ba
\vec{\rho}_1 &=& \frac{1}{\sqrt{2}} 
( \vec{r}_1 - \vec{r}_2 ) ~,
\nonumber\\
\vec{\rho}_2 &=& \frac{1}{\sqrt{6}} 
( \vec{r}_1 + \vec{r}_2 - 2\vec{r}_3 ) ~,
\nonumber\\
\vec{\rho}_3 &=& \frac{1}{\sqrt{12}} 
( \vec{r}_1 + \vec{r}_2 + \vec{r}_3 - 3\vec{r}_4 ) ~,
\nonumber\\
\vec{\rho}_4 &=& \frac{1}{\sqrt{20}} 
( \vec{r}_1 + \vec{r}_2 + \vec{r}_3 + \vec{r}_4 - 4\vec{r}_5 ) ~, 
\label{jacobi}
\ea
where $\vec{r}_i$ ($i=1,..,4$) denote the coordinate of the $i$-th quark,  
and $\vec{r}_5$ that of the antiquark. The last Jacobi coordinate 
is symmetric under the interchange of the quark coordinates, 
and hence transforms as $A_1$ under ${\cal T}_d$ ($\sim S_4$), whereas 
the first three transform as three components of $F_2$ \cite{KM}. 

Since the color part of the pentaquark wave function is a $[222]$ singlet 
and that of the antiquark a $[11]$ anti-triplet, the color wave function of 
the four-quark configuration is a $[211]$ triplet with $F_1$ symmetry.  
The total $q^4$ wave function is antisymmetric ($A_2$), hence the 
orbital-spin-flavor part has to have $F_2$ symmetry
\ba
\psi \;=\; \left[ \psi^{\rm c}_{F_1} \times 
\psi^{\rm osf}_{F_2} \right]_{A_2} ~.
\label{wf}
\ea
Here the square brackets $[\cdots]$ denote the tensor coupling under the 
tetrahedral group ${\cal T}_d$. The exotic spin-flavor states associated 
with the $S$-wave state $L^p_t=0^+_{A_1}$ belong to the 
$[f]_t=[42111]_{F_2}$ spin-flavor multiplet \cite{BGS1}. 
The corresponding orbital-spin-flavor wave function is given by 
\ba
\psi^{\rm osf}_{F_2} \;=\; 
\left[ \psi^{\rm o}_{A_1} \times \psi^{\rm sf}_{F_2} \right]_{F_2} ~, 
\label{wf1} 
\ea
where the orbital wave function depends on the Jacobi coordinates of 
Eq.~(\ref{jacobi}), $\psi^{\rm o}_t=
\psi^{\rm o}_t(\vec{\rho}_1,\vec{\rho}_2,\vec{\rho}_3,\vec{\rho}_4)$. 
A $P$-wave radial excitation with $L^p_t=1^-_{F_2}$ gives rise to exotic 
pentaquark states of the $[51111]_{A_1}$, $[42111]_{F_2}$, $[33111]_E$ and 
$[32211]_{F_1}$ spin-flavor configurations. They are characterized by 
the orbital-spin-flavor wave functions 
\ba
\psi^{\rm osf}_{F_2} \;=\; 
\left[ \psi^{\rm o}_{F_2} \times \psi^{\rm sf}_{t} \right]_{F_2} ~,
\label{wf2}
\ea
with $t=A_1$, $F_2$, $E$ and $F_1$, respectively. 
In this Letter, we study the magnetic moments of the lowest pentaquark 
antidecuplet with positive and negative parity. 

\section{Magnetic moments}

A compilation of theoretical values of the magnetic moments of exotic 
pentaquarks has been presented in \cite{Liu,Li} for the chiral soliton 
model, different correlated quark models, the MIT bag model and for QCD 
sum rules. To the best of our knowledge, the present calculation is the 
first one for an uncorrelated or constituent quark model. 
The magnetic moment of a multiquark system is given by the 
sum of the magnetic moments of its constituent parts 
\ba
\vec{\mu} \;=\; \vec{\mu}_{\rm spin} + \vec{\mu}_{\rm orb} \;=\; 
\sum_i \mu_i (2\vec{s}_{i} + \vec{\ell}_{i}) ~, 
\ea
where $\mu_i=e_i/2m_i$, $e_i$ and $m_i$ represent the magnetic moment, 
the electric charge and the constituent mass of the $i$-th (anti)quark. 
The quark magnetic moments $\mu_u$, $\mu_d$ and $\mu_s$ are determined 
from the proton, neutron and $\Lambda$ magnetic moments to be $\mu_u=1.852$ 
$\mu_N$, $\mu_d=-0.972$ $\mu_N$ and $\mu_s=-0.613$ $\mu_N$ \cite{PDG}. 
The magnetic moments of the antiquarks satisfy $\mu_{\bar{q}}=-\mu_q$. 

\subsection{Negative parity}

We first analyze the negative parity antidecuplet states that are 
associated with the $S$-wave state $L^p_t=0^+_{A_1}$ and belong to the 
$[f]_t=[42111]_{F_2}$ spin-flavor multiplet. 
The corresponding pentaquark wave function with angular momentum $J$  
and projection $M=J$ is given by Eqs.~(\ref{wf}) and~(\ref{wf1})
\ba
\psi_J &=& \frac{1}{\sqrt{3}} \left[ \psi^{\rm o}_{A_1} \left( 
\psi^{\rm c}_{F_{1\lambda}} \psi^{\rm sf}_{F_{2\rho}} 
- \psi^{\rm c}_{F_{1\rho}} \psi^{\rm sf}_{F_{2\lambda}} 
+ \psi^{\rm c}_{F_{1\eta}} \psi^{\rm sf}_{F_{2\eta}} \right) 
\right]^{(J)}_{J} ~. 
\label{wfneg}
\ea
The spin-flavor part can be expressed as a product of the antidecuplet 
flavor wave function $\phi_E$ and the $s=1/2$ spin wave function 
$\chi_{F_2}$ 
\ba
\psi^{\rm sf}_{F_{2\rho}} &=& -\frac{1}{2} \phi_{E_{\rho}} \chi_{F_{2\lambda}} 
-\frac{1}{2} \phi_{E_{\lambda}} \chi_{F_{2\rho}} 
+\frac{1}{\sqrt{2}} \phi_{E_{\rho}} \chi_{F_{2\eta}} ~,
\nonumber\\
\psi^{\rm sf}_{F_{2\lambda}} &=& -\frac{1}{2} \phi_{E_{\rho}} \chi_{F_{2\rho}} 
+\frac{1}{2} \phi_{E_{\lambda}} \chi_{F_{2\lambda}} 
+\frac{1}{\sqrt{2}} \phi_{E_{\lambda}} \chi_{F_{2\eta}} ~,
\nonumber\\
\psi^{\rm sf}_{F_{2\eta}} &=& 
 \frac{1}{\sqrt{2}} \phi_{E_{\rho}} \chi_{F_{2\rho}} 
+\frac{1}{\sqrt{2}} \phi_{E_{\lambda}} \chi_{F_{2\lambda}} ~. 
\label{wfsf}
\ea
The coefficients in Eqs.~(\ref{wfneg}) and~(\ref{wfsf}) are a consequence 
of the tensor couplings under the tetrahedral group ${\cal T}_d$ 
(Clebsch-Gordon coefficients). The total angular momentum is $J=1/2$. 
The explicit form of the spin and flavor wave functions is given in the 
appendix. Since the orbital wave function has $L^p_t=0^+_{A_1}$, the 
magnetic moment only depends on the spin part. For the $\Theta^+$, 
$\Xi_{3/2}^+$ and $\Xi_{3/2}^{--}$ exotic states we obtain 
\ba
\mu_{\Theta^+} &=& \frac{1}{3}(2\mu_u + 2\mu_d + \mu_s) 
\;=\; 0.38 \; \mu_N ~,
\nonumber\\
\mu_{\Xi_{3/2}^{--}} &=& \frac{1}{3}(\mu_u + 2\mu_d + 2\mu_s) 
\;=\; -0.44 \; \mu_N ~.
\nonumber\\
\mu_{\Xi_{3/2}^+}    &=& \frac{1}{3}(2\mu_u + \mu_d + 2\mu_s)
\;=\; 0.50 \; \mu_N ~,
\label{mmneg}
\ea
in agreement with the results obtained \cite{Liu} for the MIT bag model 
\cite{DS}. We note, that these results are independent of the orbital 
wave functions, and are valid for any quark model in which the eigenstates 
have good $SU_{\rm sf}(6)$ spin-flavor symmetry. 

\subsection{Positive parity}

Next, we study the case of positive parity antidecuplet states. These 
pentaquark states correspond to a $P$-wave state $L^p_t=1^-_{F_2}$ 
and belong to the $[f]_t=[51111]_{A_1}$ spin-flavor multiplet. 
The corresponding pentaquark wave function with 
angular momentum $J$ and projection $M=J$ is given by 
Eqs.~(\ref{wf}) and~(\ref{wf2}) 
\ba
\psi_J &=& \frac{1}{\sqrt{3}} \left[ \left( 
\psi^{\rm o}_{F_{2\rho}} \psi^{\rm c}_{F_{1\lambda}} 
- \psi^{\rm o}_{F_{2\lambda}} \psi^{\rm c}_{F_{1\rho}} 
+ \psi^{\rm o}_{F_{2\eta}} \psi^{\rm c}_{F_{1\eta}} \right) 
\psi^{\rm sf}_{A_1} \right]^{(J)}_{J} ~. 
\label{wfpos}
\ea
The spin-flavor part is now a product of the antidecuplet flavor 
wave function $\phi_E$ and the $s=1/2$ spin wave function $\chi_E$ 
\ba
\psi^{\rm sf}_{A_1} &=& \frac{1}{\sqrt{2}} \left( \phi_{E_{\rho}} 
\chi_{E_{\rho}} + \phi_{E_{\lambda}} \chi_{E_{\lambda}} \right) ~. 
\ea
The total angular momentum is $J=1/2$, $3/2$. 
The explicit form of the spin and flavor wave functions is given in the 
appendix. Since the spin of the four-quark system is $s=0$, the spin part 
of the magnetic moment only depends on the contribution from the antiquark. 
For $\Theta^+$ state with $J^p=1/2^+$ we obtain
\ba
\left< \psi_J \right| \vec{\mu}_{\rm spin} \left| \psi_J \right> \;=\; \left[ 
 \left< 1,0,\frac{1}{2},\frac{1}{2} \right| \frac{1}{2},\frac{1}{2} \right>^2
-\left< 1,1,\frac{1}{2},-\frac{1}{2} \left| \frac{1}{2},\frac{1}{2} \right>^2 
\right] \mu_{\bar{s}}  
\;=\; \frac{1}{3} \mu_s ~.
\ea
In contrast to the previous case of a negative parity pentaquark, 
now we also have a contribution from the orbital angular momentum. 
The orbital excitation with $L^p_t=1^-_{F_2}$ is an excitation in the 
relative coordinates of the four-quark subsystem. There is no excitation 
in the relative coordinate between the four-quark system and the 
antiquark. Therefore, the orbital part to the 
magnetic moment is given by  
\ba
\left< \psi_J \right| \vec{\mu}_{\rm orb} \left| \psi_J \right> 
&=& \left< \psi_J \right| \mu_2 \, \vec{\ell}_{\rho_1} 
+ \mu_3 \, \vec{\ell}_{\rho_2} + \mu_4 \, \vec{\ell}_{\rho_3} 
\left| \psi_J \right> 
\nonumber\\
&=& 3 \left< \psi_J \right| \mu_4 \, \vec{\ell}_{\rho_3} \left| \psi_J \right> 
\nonumber\\
&=& \frac{1}{2}  
\left< 1,1,\frac{1}{2},-\frac{1}{2} \left| \frac{1}{2},\frac{1}{2} \right>^2 
\right. (\mu_u + \mu_d)  
\nonumber\\
&=& \frac{1}{3}(\mu_u + \mu_d) ~. 
\ea
The total magnetic moment of the $\Theta^+$ state is 
\ba
\mu_{\Theta^+} \;=\;  
\frac{1}{3}(\mu_u + \mu_d + \mu_s) \;=\; 0.09 \; \mu_N ~. 
\ea
The magnetic moments of the exotic pentaquarks of the antidecuplet 
$\Theta^+$, $\Xi_{3/2}^+$ and $\Xi_{3/2}^{--}$ with angular momentum 
and parity $J^p=1/2^+$ are equal \footnote{In this calculation we have 
used harmonic oscillator wave functions with $N=1$.}  
\ba
\mu_{\Xi_{3/2}^{--}} \;=\; \mu_{\Xi_{3/2}^+} \;=\; \mu_{\Theta^+} 
\;=\; \frac{1}{3}(\mu_u + \mu_d + \mu_s) \;=\; 0.09 \; \mu_N ~.
\label{mmpos1}
\ea
The only difference for pentaquarks with angular momentum and parity 
$J^p=3/2^+$ is in the angular momentum couplings. As a result we find 
\ba
\mu_{\Theta^+} &=& \frac{1}{2}(\mu_u + \mu_d - 2\mu_s) 
\;=\; 1.05 \; \mu_N ~,
\nonumber\\
\mu_{\Xi_{3/2}^{--}} &=& \frac{1}{2}(- 2\mu_u + \mu_d + \mu_s)
\;=\; -2.64 \; \mu_N ~,
\nonumber\\
\mu_{\Xi_{3/2}^+}    &=& \frac{1}{2}(\mu_u - 2\mu_d + \mu_s)
\;=\; 1.59 \; \mu_N ~.
\label{mmpos2}
\ea

\subsection{Sum rules}

The results obtained for the magnetic moments are valid for any constituent 
quark model in which the eigenstates have good $SU_{\rm sf}(6)$ spin-flavor 
symmetry. In Table~\ref{mmpenta}, we present the magnetic moments of all 
antidecuplet pentaquarks for the three different combinations of 
angular momentum and parity discussed in the previous section, i.e. 
$J^p=1/2^-$, $1/2^+$ and $3/2^+$. In all three cases, the magnetic moments 
satisfy the generalized Coleman-Glashow sum rules \cite{coleman,hong} 
\ba
\mu_{\Theta^+} + \mu_{\Xi^+_{3/2}} &=& \mu_{N^+} + \mu_{\Sigma^+} ~,
\nonumber\\
\mu_{\Theta^+} + \mu_{\Xi^{--}_{3/2}} &=& \mu_{N^0} + \mu_{\Sigma^-} ~, 
\nonumber\\
\mu_{\Xi^{--}_{3/2}} + \mu_{\Xi^+_{3/2}} &=& \mu_{\Xi^-_{3/2}} + 
\mu_{\Xi^0_{3/2}} ~,
\ea
and
\ba
2\mu_{\Sigma^0} \;=\; \mu_{\Sigma^-} + \mu_{\Sigma^+} 
\;=\; \mu_{N^0} + \mu_{\Xi^0_{3/2}} 
\;=\; \mu_{N^+} + \mu_{\Xi^-_{3/2}} ~. 
\ea
The same sum rules hold for the chiral quark-soliton model 
in the chiral limit \cite{Kim1}. 
In addition, there exist interesting sum rules that relate the magnetic 
moments of the antidecuplet pentaquarks to those of the decuplet 
and octet baryons. For the case of negative parity pentaquarks, 
we use Eq.~(\ref{mmneg}) to obtain the sum rules 
\ba
\mu_{\Theta^+} - \mu_{\Xi_{3/2}^{--}} 
&=& \frac{1}{9} \left( \mu_{\Delta^{++}} - \mu_{\Omega^-} \right) 
\;=\; \frac{1}{12} \left( 2\mu_p + \mu_n + \mu_{\Sigma^+} - \mu_{\Sigma^-} 
- \mu_{\Xi^0} - 2\mu_{\Xi^-} \right) ~,
\nonumber\\
\mu_{\Theta^+} - \mu_{\Xi_{3/2}^+} 
&=& \frac{1}{9} \left( \mu_{\Delta^-} - \mu_{\Omega^-} \right) 
\;=\; \frac{1}{12} \left( \mu_p + 2\mu_n - \mu_{\Sigma^+} + \mu_{\Sigma^-} 
- 2\mu_{\Xi^0} - \mu_{\Xi^-} \right) ~,
\nonumber\\
\mu_{\Xi_{3/2}^+} - \mu_{\Xi_{3/2}^{--}} 
&=& \frac{1}{9} \left( \mu_{\Delta^{++}} - \mu_{\Delta^-} \right) 
\;=\; \frac{1}{12} \left( \mu_p - \mu_n + 2\mu_{\Sigma^+} - 2\mu_{\Sigma^-} 
+ \mu_{\Xi^0} - \mu_{\Xi^-} \right) ~.
\label{sumrule1}
\ea
The first sum rule is similar, but not identical, to the result obtained 
in the chiral quark-soliton model \cite{Kim1,Kim2}. 
For the positive parity pentaquarks, the results of Eq.~(\ref{mmpos2}) 
can be used to obtain the sum rules that relate the 
magnetic moments of the $J^p=3/2^+$ antidecuplet pentaquarks to those of 
the decuplet baryons
\ba
\mu_{\Theta^+} - \mu_{\Xi_{3/2}^{--}} 
&=& \frac{9}{2} \left( \mu_{\Delta^{++}} - \mu_{\Omega^-} \right) ~,
\nonumber\\
\mu_{\Theta^+} - \mu_{\Xi_{3/2}^+} 
&=& \frac{9}{2} \left( \mu_{\Delta^-} - \mu_{\Omega^-} \right) ~,
\nonumber\\
\mu_{\Xi_{3/2}^+} - \mu_{\Xi_{3/2}^{--}} 
&=& \frac{9}{2} \left( \mu_{\Delta^{++}} - \mu_{\Delta^-} \right) ~.
\label{sumrule2}
\ea

In the limit of equal quark masses $m_u=m_d=m_s=m$, the magnetic moments 
of the antidecuplet pentaquark states (denoted by $i \in \overline{10}$) 
become proportional to the electric charges 
\ba
\mu_i \;=\; \frac{1}{9} \frac{1}{2m} Q_i ~,
&\hspace{1cm}& \mbox{for } J^p=1/2^- ~,
\nonumber\\
\mu_i \;=\; \frac{1}{2} \frac{1}{2m} Q_i ~,
&\hspace{2cm}&  \mbox{for } J^p=3/2^+ ~,
\label{mmq}
\ea
compared to 
\ba
\mu_i \;=\; \frac{1}{2m} Q_i ~, 
\ea
for the decuplet baryons ($i \in 10$). 
For the exotic pentaquarks, Eq.~(\ref{mmq}) 
implies $\mu_{\Xi_{3/2}^{--}}= -2 \mu_{\Xi_{3/2}^+} = -2 \mu_{\Theta^+}$. 
For angular momentum and parity $J^p=1/2^+$, the magnetic moments vanish 
in the limit of equal quark masses due to a cancellation between the spin 
and orbital contributions. For all three cases, the sum of the magnetic 
moments of all members of the antidecuplet vanishes identically 
\ba
\sum_{i \in \overline{10}} \mu_i \;=\; 0 ~.
\ea

\section{Discussion}

The magnetic moments for negative parity $J^p=1/2^-$ pentaquarks 
of Eq.~(\ref{mmneg}) are typically an order of magnitude smaller than the 
proton magnetic moment, whereas for positive parity $J^p=1/2^+$ they are 
even smaller due to a cancellation between orbital and spin contributions, 
see Eq.~(\ref{mmpos1}). The largest values of the magnetic moment are  
obtained for $J^p=3/2^+$ pentaquarks, but they are still smaller than the 
proton value. The magnetic moment of the $\Theta(1540)$ in the constituent 
quark model is found to be 0.38, 0.09 and 1.05 $\mu_N$ for $J^p=1/2^-$, 
$1/2^+$ and $3/2^+$, respectively. 
In Tables~\ref{mmnegpar} and~\ref{mmpospar} we present a comparison with 
other theoretical predictions for the magnetic moments of exotic pentaquarks 
with negative and positive parity, respectively. 

The first estimate of the $\Theta^+$ magnetic moment was made by Nam, 
Hosaka and Kim in a study of photoproduction reactions \cite{Nam}. They 
used the diquark model of \cite{JW} (JW) to estimate the anomalous magnetic 
moment as $\kappa=-0.7$ for positive and $-0.2$ for negative parity. 
For the $\Theta^+$ as a $KN$ bound state, they obtained $\kappa=-0.4$ for 
positive and $-0.5$ for negative parity. In all cases, the spin is $J=1/2$. 

In \cite{Kim1}, Kim and Prasza{\l}owicz investigated the magnetic moments 
of the baryon antidecuplet in the chiral soliton model in the chiral limit 
($\chi$QCD). 
The spin and parity are $J^p=1/2^+$. The $\Theta^+$ magnetic moment was 
found to be 0.12, 0.20 or 0.30 $\mu_N$, depending on three different ways 
to determine the parameters. The magnetic moments of the exotic cascade 
pentaquarks are obtained from the proportionality of the magnetic 
moments of the antidecuplet baryons to the electric charge. 
We note that in this calculation no $SU(3)$ symmetry breaking effects were 
taken into account, unlike the other approaches discussed in this section. 

Also Zhao used the diquark model of \cite{JW} to obtain an anomalous 
magnetic moment $\kappa=-0.87$ for a positive parity $\Theta^+$ pentaquark 
\cite{Zhao}. For the case of negative parity, the magnetic moment was 
estimated from the sum of $u\bar{s}$ and $udd$ clusters to be  
0.60 $\mu_N$.

Huang et al. used light cone QCD sum rules to extract the absolute value of 
the $\Theta(1540)$ magnetic moment as $0.12 \pm 0.06$ $\mu_N$ 
\cite{Huang}. In this calculation, the $\Theta^+$ was assumed to be an 
isoscalar with spin $J=1/2$, no assumption was made of its parity. 

Finally, there are two studies in which the magnetic moments of exotic 
antidecuplet baryons are calculated for spin $J=1/2$ \cite{Liu} and $J=3/2$ 
\cite{Li} for a variety of models of pentaquarks: 
the diquark-diquark-antiquark models of \cite{JW} and \cite{SZ} (SZ), 
as a diquark-triquark bound state \cite{KL} (KL), and the MIT bag model 
\cite{DS}. In the latter case, the parity is negative, whereas in 
all others it is positive. 

Although there is some variation in the numerical values obtained 
for different models of pentaquarks, generally speaking, the predictions 
for the magnetic moment of the $\Theta^+$ are relatively close, especially 
in comparison with the magnetic moment of the proton they are all small. 
For the case of negative parity pentaquarks, our results are identical 
to those derived for the MIT bag model \cite{Liu}. The small difference 
in the numerical values is due to the values of the quark magnetic 
moments $\mu_q$ used in the calculations. For positive parity pentaquarks 
with $J^p=1/2^+$ the values of the magnetic moments are suppressed due to 
cancellations between the spin and orbital contributions. 
Our predictions for the magnetic moments of the exotic pentaquarks with 
$J^p=3/2^+$ are in qualitative agreement with those of the diquark 
models of \cite{JW} and \cite{SZ}, but differ somewhat from the ones 
for the diquark-triquark cluster model of \cite{KL}. 

\section{Summary and conclusions}

In this Letter, we have analyzed the pentaquark magnetic moments of the 
lowest flavor antidecuplet for both positive and negative parity in the 
constituent quark model. The resulting magnetic moments 
were obtained in closed analytic form, which made it possible to derive 
generalized Coleman-Glashow sum rules for the antidecuplet magnetic 
moments, as well as sum rules connecting the magnetic moments of 
antidecuplet pentaquarks to those of decuplet and octet baryons. 
The numerical values are in qualitative agreement with those obtained 
in other approaches, such as correlated quark models, QCD sum rules, 
MIT bag model and the chiral soliton model.  

In conclusion, the spectroscopy of exotic baryons will be a key testing 
ground for models of baryons and their structure. Especially the measurement 
of the angular momentum and parity of the $\Theta^+(1540)$ may help to 
distinguish between different models and to gain more insight into 
the relevant degrees of freedom and the underlying dynamics that determines 
the properties of exotic baryons. The magnetic moment is an important 
ingredient for the calculation of the total and differential cross sections 
for photo- and electroproduction which have been proposed as a tool to 
help determine the quantum numbers of the $\Theta^+$ pentaquarks. 
The values of the magnetic moments presented here, together with those 
of \cite{Liu,Li}, may be used as an input for such calculations. 

\section*{Acknowledgements}

This work is supported in part by a grant from CONACyT, M\'exico. 

\appendix

\section{Spin wave functions}

The spin wave function with $s=1/2$ and $F_2$ symmetry is a combination 
of the spin wave function for the four-quark system with $[31]$ and $s=1$ 
and that of the antiquark with $s=1/2$. We start by decoupling the spin 
of the antiquark
\ba
\chi_{F_{2\alpha}} &=& 
\left| [31],1,\frac{1}{2};\frac{1}{2},\frac{1}{2} \right>_{F_{2\alpha}}
\nonumber\\
&=& \sqrt{\frac{2}{3}} \, 
\left| [31],1,1 \right>_{F_{2\alpha}} \downarrow 
-\sqrt{\frac{1}{3}} \,
\left| [31],1,0 \right>_{F_{2\alpha}} \uparrow ~, 
\ea
with $\alpha=\rho$, $\lambda$, $\eta$. The $\uparrow$ and $\downarrow$ 
represent the spin of the antiquark. 
The spin wave functions of the four-quark system are given by
\ba
\left| [31],1,1 \, \right>_{F_{2\rho}} &=& 
-\frac{1}{\sqrt{2}} \; \left| \; \downarrow \uparrow \uparrow \uparrow 
- \uparrow \downarrow \uparrow \uparrow \; \right> ~, 
\nonumber\\
\left| [31],1,1 \, \right>_{F_{2\lambda}} &=& 
-\frac{1}{\sqrt{6}} \; \left| \; 
   \downarrow \uparrow \uparrow \uparrow 
+  \uparrow \downarrow \uparrow \uparrow 
-2 \uparrow \uparrow \downarrow \uparrow \; \right> ~,
\nonumber\\
\left| [31],1,1 \, \right>_{F_{2\eta}} &=& 
-\frac{1}{2\sqrt{3}} \; \left| \;
   \downarrow \uparrow \uparrow \uparrow 
+  \uparrow \downarrow \uparrow \uparrow 
+  \uparrow \uparrow \downarrow \uparrow 
-3 \uparrow \uparrow \uparrow \downarrow \; \right> ~.
\ea
The states with other values of the projection $m_s$ can be obtained 
by applying the lowering operator in spin space. 

The spin wave function with $s=1/2$ and $E$ symmetry is a combination 
of the spin wave function for the four-quark system with $[22]$ and $s=0$  
and that of the antiquark with $s=1/2$ 
\ba
\chi_{E_{\alpha}} \;=\; 
\left| [22],0,\frac{1}{2};\frac{1}{2},\frac{1}{2} \right>_{E_{\alpha}}
\;=\; \left| [22],0,0 \right>_{E_{\alpha}} \uparrow ~, 
\ea
with $\alpha=\rho$, $\lambda$. In this case, the spin wave functions 
of the four-quark system are given by
\ba
\left| [22],0,0 \, \right>_{E_{\rho}} &=& 
-\frac{1}{2} \; \left| \; 
   \downarrow \uparrow \uparrow \downarrow 
-  \uparrow \downarrow \uparrow \downarrow 
+  \uparrow \downarrow \downarrow \uparrow 
-  \downarrow \uparrow \downarrow \uparrow \; \right> ~,
\nonumber\\
\left| [22],0,0 \, \right>_{E_{\lambda}} &=& 
-\frac{1}{2\sqrt{3}} \; 
\left| \; 
   \downarrow \uparrow \uparrow \downarrow 
+  \uparrow \downarrow \uparrow \downarrow 
-2 \uparrow \uparrow \downarrow \downarrow 
+  \uparrow \downarrow \downarrow \uparrow 
+  \downarrow \uparrow \downarrow \uparrow 
-2 \downarrow \downarrow \uparrow \uparrow \; \right> ~. 
\ea

\section{Flavor wave functions}

The flavor wave functions for the antidecuplet $\Theta^+$ pentaquark 
with $I=I_3=0$ are given by 
\ba
\phi_{E_{\rho}} &=& -\frac{1}{2} (duud-udud+uddu-dudu)\bar{s} ~, 
\nonumber\\
\phi_{E_{\lambda}} &=& -\frac{1}{2\sqrt{3}} 
(duud+udud-2uudd+uddu+dudu-2dduu)\bar{s} ~.
\ea
The flavor states with other values of the isospin $I$, its projection 
$I_3$ and hypercharge $Y$ can be obtained by applying the ladder operators 
in flavor space and using the phase convention of De Swart \cite{deSwart}.

\clearpage

\begin{table}
\centering
\caption[Antidecuplet magnetic moments]{Magnetic moments of 
antidecuplet pentaquarks in $\mu_N$ for 
$J^p=\frac{1}{2}^-$, $\frac{1}{2}^+$ and $\frac{3}{2}^+$.}
\label{mmpenta}
\vspace{15pt}
\begin{tabular}{lccc}
\hline
& & & \\
& $J^p=\frac{1}{2}^-$ & $J^p=\frac{1}{2}^+$ & $J^p=\frac{3}{2}^+$ \\
& & & \\
\hline
& & & \\
$\Theta^+$ & $\frac{1}{9}(6\mu_u + 6\mu_d + 3\mu_s)$ 
& $\frac{1}{3}(\mu_u + \mu_d + \mu_s)$
& $\frac{1}{2}(\mu_u + \mu_d - 2\mu_s)$ \\
$N^0$      & $\frac{1}{9}(5\mu_u + 6\mu_d + 4\mu_s)$ 
& $\frac{1}{3}(\mu_u + \mu_d + \mu_s)$
& $\frac{1}{2}(\mu_d - \mu_s)$ \\
$N^+$      & $\frac{1}{9}(6\mu_u + 5\mu_d + 4\mu_s)$ 
& $\frac{1}{3}(\mu_u + \mu_d + \mu_s)$
& $\frac{1}{2}(\mu_u - \mu_s)$ \\
$\Sigma^-$ & $\frac{1}{9}(4\mu_u + 6\mu_d + 5\mu_s)$ 
& $\frac{1}{3}(\mu_u + \mu_d + \mu_s)$
& $\frac{1}{2}(- \mu_u + \mu_d)$ \\
$\Sigma^0$ & $\frac{1}{9}(5\mu_u + 5\mu_d + 5\mu_s)$ 
& $\frac{1}{3}(\mu_u + \mu_d + \mu_s)$
& $0$ \\
$\Sigma^+$ & $\frac{1}{9}(6\mu_u + 4\mu_d + 5\mu_s)$ 
& $\frac{1}{3}(\mu_u + \mu_d + \mu_s)$
& $\frac{1}{2}(\mu_u - \mu_d)$ \\
$\Xi^{--}_{3/2}$ & $\frac{1}{9}(3\mu_u + 6\mu_d + 6\mu_s)$ 
& $\frac{1}{3}(\mu_u + \mu_d + \mu_s)$
& $\frac{1}{2}(- 2\mu_u + \mu_d + \mu_s)$ \\
$\Xi^{-}_{3/2}$  & $\frac{1}{9}(4\mu_u + 5\mu_d + 6\mu_s)$ 
& $\frac{1}{3}(\mu_u + \mu_d + \mu_s)$
& $\frac{1}{2}(- \mu_u + \mu_s)$ \\
$\Xi^{0}_{3/2}$  & $\frac{1}{9}(5\mu_u + 4\mu_d + 6\mu_s)$ 
& $\frac{1}{3}(\mu_u + \mu_d + \mu_s)$
& $\frac{1}{2}(- \mu_d + \mu_s)$ \\
$\Xi^{+}_{3/2}$  & $\frac{1}{9}(6\mu_u + 3\mu_d + 6\mu_s)$ 
& $\frac{1}{3}(\mu_u + \mu_d + \mu_s)$
& $\frac{1}{2}(\mu_u - 2\mu_d + \mu_s)$ \\
& & & \\
\hline
& & & \\
$\sum_{i \in \overline{10}} \mu_i$ & $\frac{50}{9}(\mu_u+\mu_d+\mu_s)$ 
& $\frac{10}{3}(\mu_u+\mu_d+\mu_s)$ & 0 \\
& & & \\
\hline
\end{tabular}
\end{table}

\begin{table}[ht]
\centering
\caption[]{Comparison of magnetic moments in $\mu_N$ of exotic 
antidecuplet pentaquarks with angular momentum and parity 
$J^p=\frac{1}{2}^-$.}
\label{mmnegpar}
\vspace{15pt}
\begin{tabular}{lccrr}
\hline
& & & & \\
& & \multicolumn{3}{c}{$J^p=\frac{1}{2}^-$}\\
Method & Ref. & $\Theta^+$ & $\Xi_{3/2}^+$ & $\Xi_{3/2}^{--}$ \\
& & & & \\
\hline
& & & & \\
Present & & 0.38 & 0.50 & --0.44 \\
& & & & \\
MIT bag & \protect{\cite{Liu}} & 0.37 & 0.45 & --0.42 \\
JW diquark & \protect{\cite{Nam}}    & 0.49 & & \\
$KN$ bound state & \protect{\cite{Nam}}    & 0.31 & & \\
Cluster & \protect{\cite{Zhao}}   & 0.60 & & \\
QCD sum rules & \protect{\cite{Huang}}  & $0.12 \pm 0.06^{\ast}$ & & \\
& & & & \\
\hline
& & & & \\
\multicolumn{5}{l}{$^{\ast}$ Absolute value}
\end{tabular}
\end{table}

\begin{table}[ht]
\centering
\caption[]{Comparison of magnetic moments in $\mu_N$ of exotic 
antidecuplet pentaquarks with angular momentum and parity 
$J^p=\frac{1}{2}^+$ and $\frac{3}{2}^+$.}
\label{mmpospar}
\vspace{15pt}
\begin{tabular}{lccrrcrrr}
\hline
& & & & & & & & \\
& & \multicolumn{3}{c}{$J^p=\frac{1}{2}^+$} & 
& \multicolumn{3}{c}{$J^p=\frac{3}{2}^+$} \\
Method & Ref. & $\Theta^+$ & $\Xi_{3/2}^+$ & $\Xi_{3/2}^{--}$ & 
Ref. & $\Theta^+$ & $\Xi_{3/2}^+$ & $\Xi_{3/2}^{--}$ \\
& & & & & & & & \\
\hline
& & & & & & & & \\
Present & & 0.09 & 0.09 & 0.09 & & 1.05 & 1.59 & --2.64 \\
& & & & & & & & \\
JW diquark & \protect{\cite{Zhao,Liu}} & 0.08 & --0.06 &   0.12 & 
\protect{\cite{Li}}       & 1.01 &   1.22 & --2.43 \\
SZ diquark & \protect{\cite{Liu}}      & 0.23 &   0.33 & --0.17 & 
\protect{\cite{Li}}       & 1.23 &   1.85 & --2.84 \\
KL cluster & \protect{\cite{Liu}}      & 0.19 &   0.13 & --0.43 & 
\protect{\cite{Li}}       & 0.84 &   0.89 & --1.20 \\
$\chi$QSM & \protect{\cite{Kim1}} & 0.12 & 0.12 & --0.24 & & & & \\
$\chi$QSM & \protect{\cite{Kim1}} & 0.20 & 0.20 & --0.40 & & & & \\
$\chi$QSM & \protect{\cite{Kim1}} & 0.30 & 0.30 & --0.60 & & & & \\
JW diquark & \protect{\cite{Nam}} & 0.18 & & & & & & \\
$KN$ bound state & \protect{\cite{Nam}} & 0.36 & & & & & \\
QCD sum rules & \protect{\cite{Huang}}& $0.12 \pm 0.06^{\ast}$ & & & & & \\
& & & & & & & \\
\hline
& & & \\
\multicolumn{9}{l}{$^{\ast}$ Absolute value}
\end{tabular}
\end{table}


\begin{thebibliography}{99}
\bibitem{Kim2}
Note that in Eq.~(19) of Ref.~\cite{Kim1} a term $-3v_2/8$ is missing. 
The present result of Eq.~(\ref{sumrule1}) is obtained for $v_2=40(w-v)/3$ 
which, for the three fits considered in \cite{Kim1}, would correspond to 
numerical values of $v_2=4.91$, 4.37 and 4.00, respectively, quite close 
to the value of $v_2=5$ used in the calculation of the magnetic moments 
in \cite{Kim1}.
\end{thebibliography}
\end{document}